# Dimensional Control of Octahedral Tilt in SrRuO$_3$ via Infinite-layered Oxides

Shan Lin,[†,‡] Qinghua Zhang,[†] Xiahan Sang,[§] Jiali Zhao,[†] Sheng Cheng,[⊥] Amanda Huon,[▽] Qiao Jin,[†,#] Shuang Chen,[†,○] Shengru Chen,[†,#] Haizhong Guo,[○] Meng He,[†] Chen Ge,[†] Can Wang,[†,#,♦] Jia-Ou Wang,[l] Michael R. Fitzsimmons,[▽] Lin Gu,[†,#,♦] Tao Zhu,[†,#,⊥,♦] Kui-juan Jin,[*,†,#,♦] and Er-Jia Guo[*,†,‡,#,♦]

[†] Beijing National Laboratory for Condensed Matter Physics and Institute of Physics, Chinese Academy of Sciences, Beijing 100190, China

[‡] Center of Materials Science and Optoelectronics Engineering, University of Chinese Academy of Sciences, Beijing 100049, China

[§] State Key Laboratory of Advanced Technology for Materials Synthesis and Processing and Nanostructure research center, Wuhan University of Technology, 122 Luoshi Rd., Wuhan 430070, China

[l] Institute of High Energy Physics, Chinese Academy of Sciences, Beijing 100049, China

[⊥] Spallation Neutron Source Science Center, Dongguan 523803, China

[#] School of Physical Sciences, University of Chinese Academy of Sciences, Beijing 100190, China

▽Neutron Scattering Division, Oak Ridge National Laboratory, Oak Ridge, TN 37831, United States

° School of Physical Engineering, Zhengzhou University, Zhengzhou 450001, China

♦ Songshan Lake Materials Laboratory, Dongguan, Guangdong 523808, China

**KEYWORDS:** low-dimensional heterostructure; magnetic anisotropy; anomalous Hall conductance; polarized neutron reflectometry

**ABSTRACT**

Manipulation of octahedral distortion at atomic length scale is an effective means to tune the physical ground states of functional oxides. Previous work demonstrates that epitaxial strain and film thickness are variable parameters to modify the octahedral rotation and tilt. However, selective control of bonding geometry by structural propagation from adjacent layers is rarely studied. Here we propose a new route to tune the ferromagnetic response in $SrRuO_3$ (SRO) ultrathin layers by oxygen coordination of adjacent $SrCuO_2$ (SCO) layers. The infinite-layered $CuO_2$ in SCO exhibits a structural transformation from "planar-type" to "chain-type" as reducing film thickness. These two orientations dramatically modify the polyhedral connectivity at the interface, thus altering the octahedral distortion of SRO. The local structural variation changes the spin state of Ru and hybridization strength between Ru $4d$ and O $2p$ orbitals, leading to a significant change in the magnetoresistance and anomalous Hall resistivity of SRO layers. These findings could launch further investigations into adaptive control of magnetoelectric properties in quantum oxide heterostructures using oxygen coordination.

**Main Text**

Engineering octahedral connectivity has been a central research theme in complex oxide heterostructures over the last decades.[1,2] Atomic modification of octahedral rotation and tilt dramatically changes the bond angle and bond length, resulting in an effective control of directional hybridization of *d*-electrons and orbital degeneracy. The transport and magnetic properties of correlated perovskite oxides will be modified accordingly due to the intrinsic strong electron-lattice correlations.[3,4] Conventionally, the substrate-induced epitaxial strain effectively changes the octahedral parameters, whose tilting degree and orientation are controlled by the magnitude and sign of the misfit strain.[5-7] The misfit strain relaxes with increasing film thickness. The modification of octahedra normally does not exceed tens of unit cells (u. c.) in a single film, leading to the thickness-dependent electronic and magnetic phase transitions.[8,9]

$SrRuO_3$ (SRO) is a typical ferromagnetic metal with orthorhombic crystalline symmetry (*Pbnm*) in its bulk form.[10-13] It has attracted considerable attention because the structural symmetry and octahedral distortion of SRO thin films are extremely sensitive to the type of strain. Under the compressive strain, SRO exhibits a tetragonal structure.[14] In contrast, a monoclinic or orthorhombic structure is stabilized in the tensile strained SRO films, similar to bulk SRO. Octahedral tilt propagated from the substrate preferentially occurs when SRO is under a tensile strain as opposed to a compressive strain. Previous work demonstrates that the electrical conductivity of a compressively strained SRO film is lower than that of a film under tensile strain.[15] The magnetic easy-axis of SRO changes from out-of-plane to in-plane direction when the strain state switches from compressive to tensile strain.[16] Another approach to modify the octahedral distortion of SRO is the interfacial engineering of oxygen coordination environment by inserting a buffer layer (such as, $Ca_{0.5}Sr_{0.5}TiO_3$)[17] or capping a top layer (such as, $SrTiO_3$,[18]

LaNiO$_3$,[19] *etc*) with different crystalline symmetries and octahedral tilt patterns. This leads to the stabilization of a Ru-O-Ru bond angle in the entire SRO layer that is dramatically different from an SRO single layer or bulk SRO. The octahedral distortion thus implicitly controls the magnetic anisotropy and triggers an emergent topological Hall effect.[20-23] Recent work demonstrates that the local ionic displacement at BaTiO$_3$/SRO interface could be switched electrically, leading to a nonvolatile and reversibly tunable skyrmion due to the Dzyaloshinskii–Moriya (DM) interaction and ferroelectric proximity effect.[24]

While various approaches have been applied to modify the octahedral distortion of SRO, determinative control of bonding geometry via the structural propagation from adjacent layers is the rare case. Recently, Cui *et al*.[25] and Jeong *et al*.[26] report subsequently that the structural evolution and magnetic properties of SrTiO$_3$/SRO superlattices. The cubic symmetry of SrTiO$_3$ suppresses the octahedral tilt of SRO with increasing the thickness of adjacent SrTiO$_3$ layers, leading to an emergent correlation-driven spin-orbital coupling and a rotation of magnetic easy axis. Recent works demonstrate that atomically thin SRO layers sandwiched between SrTiO$_3$ maintain a ferromagnetic state up to 25 K and the electrons are localized in the two-dimensional space.[27,28] However, so far, the structural modification of SRO has been only observed at the heterostructures consisting of all perovskite oxides. Interface effects in the SRO heterostructures with dissimilar crystalline symmetry is scarcely reported, suggesting a great potential in designing artificial heterostructures for harvesting novel functionalities.

In this work, we demonstrate the first example of using structural transformation in an infinite layered SrCuO$_2$ (SCO) compound to tune magnetic responses in an adjacent SRO ultrathin layer. The octahedral distortion and lattice tetragonality of SRO layers are controlled by the oxygen coordination of SCO, in which the CuO$_2$ layers transform from a "*planar-type*" SCO (P-SCO) to "*chain-type*" SCO (C-SCO) as reducing film thickness below 5 u.c.[29,30] We observe an enhanced magnetoresistance but reduced anomalous Hall conductance in SRO layers with suppressed octahedral tilt. This result demonstrates a methodology of atomic-precision control of structural propagation across the heterointerfaces and resultant magnetic states in correlated electronic materials.

Two sets of [SCO$_n$/SRO$_6$/SCO$_n$] (C$_n$R$_6$C$_n$) trilayers were deposited on (001)-oriented SrTiO$_3$ substrates by pulsed laser deposition, where *n* represents the number of u. c. of SCO layers. The SRO layers are sandwiched between two inactive SCO layers to reduce the impact from SrTiO$_3$ substrates, as shown in **Figure 1**a. The SCO is used as inactive layers to tune the structural parameters of SRO. We fix the layer thickness of SRO to be 6 u. c., which is just beyond the critical thickness of metal-to-insulator transition (MIT) (Supporting Information, Fig. S1). In this phase-instability regime, the magnetic and transport properties of SRO ultrathin layer are extremely sensitive to the structural variations. The small thickness of C$_n$R$_6$C$_n$ trilayers guarantees that all layers are fully strained to the substrates. X-ray reflectivity measurements indicate the smooth interfaces and as-designed thicknesses of each layer (Supporting Information, Fig. S2). Figure 1b shows a cross-sectional high-angle annular dark-field (HAADF) STEM image of a representative C$_3$R$_6$C$_3$ trilayer. STEM results indicate that our sample shows highly epitaxial growth, good crystalline quality, and atomically sharp interfaces. Annular bright field (ABF) STEM measurements were performed to illustrate the oxygen coordination of SCO in the trilayers. The SCO forms an infinite-layer structure in both trilayers. In C$_3$R$_6$C$_3$, C-SCO exhibits vertical-aligned "*chain-like*" CuO$_2$ planes with charge neutral SrO$^0$ and CuO$^0$ sublayers. As shown in Figure 1c, an oxygen atom moves from CuO$_2$ layer to the apical oxygen vacancy to keep the electrostatic stable. As increasing the SCO layer thickness, CuO$_2$ layers in C$_{12}$R$_6$C$_{12}$ transit into a planar-arranged oxygen coordination with $Sr^{2+}$ and $CuO_2^{2-}$ sublayers, resulting in P-SCO has a perovskite-like structure with apical oxygen vacancies (Figure 1e). We performed the x-ray diffraction measurements on two trilayers (Supporting Information, Fig. S2). The out-of-plane lattice constant of SCO reduces from ~3.92(1) Å to ~3.45(1) Å when the oxygen coordination changes from "*chain-type*" to "*planar-type*" structure, is consistence with previous observations.[31,32]

The structural transformation in SCO triggers the redistribution of electrons and directly increases the electronic anisotropy. We perform x-ray absorption spectra (XAS) measurements in a 6-u.c.-thick SRO single layer, $C_3R_6C_3$, and $C_{12}R_6C_{12}$ trilayers for both O *K*- and Cu *L*-edges (Supporting Information, Fig. S3). The absorption peak at ~535 eV in O *K*-edge spectra reflects the hybridization between Ru 4*d* state and O 2*p* state. The peak intensity in $C_{12}R_6C_{12}$ is larger compared to that of $C_3R_6C_3$, indicating more unoccupied states, i.e. the lower electron population, of the hybridized orbitals in $C_{12}R_6C_{12}$. The XAS at Cu *L*-edges for both trilayers confirms the Cu ion keeps a +2-valence state and does not change with oxygen coordination of SCO. In order to directly probe the electronic occupancy in the Cu *d*-orbitals, we measured the XAS using the linearly polarized x-ray beam with two different incidence angles (30º and 90º) with respect to the surface's plane. We calculate $I_c$ and $I_{ab}$ (Figures 1d and 1f) which directly reflect the unoccupied states in the out-of-plane and in-plane orbitals, respectively. The line shapes and intensities of $I_c$ and $I_{ab}$ are nearly identical for $C_3R_6C_3$, whereas those for $C_{12}R_6C_{12}$ are completely different. Direct comparison of orbital polarization in both trilayers, shown in the bottom panels of Figures 1d and 1e, is conducted by calculating the x-ray linear dichroism (XLD = $I_c - I_{ab}$) of $C_3R_6C_3$ and $C_{12}R_6C_{12}$ trilayers. XLD for $C_3R_6C_3$ is close to zero, suggesting the orbital occupancies of Cu $d_{x^2-y^2}$ and $d_{3z^2-r^2}$ orbitals are nearly equal. This result is consistent with the symmetry character of "*chain-arranged*" $CuO_2$ layer in $C_3R_6C_3$. In contrast, we observe a non-zero XLD for $C_{12}R_6C_{12}$. At the Cu *L*-edges, $I_c$ is significantly larger than $I_{ab}$, demonstrating an anisotropic orbital polarization with electrons preferentially occupying the in-plane orbitals. Meanwhile, the pre-peak around ~529 eV at the O *K*-edges reflects the hybridization strength between Cu 3*d* orbitals and O 2*p* orbitals. We find that the intensity of $I_{ab}$ is larger than that of $I_c$, which agrees with "*planar-type*" $CuO_2$ infinite-layered structure in $C_{12}R_6C_{12}$.

To evaluate how the bonding geometry of SCO influence the physical properties of adjacent SRO layer, we performed the magnetotransport measurements on both $C_3R_6C_3$ and $C_{12}R_6C_{12}$ trilayers. **Figure 2**a shows the *ρ-T* curves of a SRO single layer and two trilayers when the magnetic fields of 0 and 9 T is applied along the out-of-plane direction, which corresponds to the magnetic easy-axis of SRO ultrathin layers. In contrast to the MIT observed in a 6-u.c-thick SRO single layer, both trilayers exhibit an insulating behavior at all temperatures. At 5 K, the resistivity of $C_{12}R_6C_{12}$ is two orders of magnitude larger than that of $C_3R_6C_3$. We investigate the magnetic anisotropy in $C_3R_6C_3$ and $C_{12}R_6C_{12}$ by measuring the field-dependent magnetoresistances [MR, $(\rho_H-\rho_0)/\rho_0$]. Figures 2b and 2c show MR for both trilayers when the magnetic field is applied parallel to the out-of-plane and in-plane direction, respectively. When *H*//*c*, MR of trilayers shows negative values and symmetric peaks at small *H* that is coupled with the coercive fields ($H_C$), in agreement with the typical behavior of a ferromagnet. However, MR reduces dramatically, and the butterfly-like hysteresis loops disappear when *H*//*ab*. A series of magnetotransport measurements were conducted at various temperatures (Supporting Information, Figures S4 and S5). Figure 2d summarizes MR of $C_3R_6C_3$ and $C_{12}R_6C_{12}$ as a function of temperature when *H*//*c* and *H*//*ab*. In order to evaluate the magnetic anisotropy, we compare the anisotropic magnetoresistance (AMR, $MR_c - MR_{ab}$) of two trilayers, as shown in Figure 2e. Comparing to AMR of a SRO single layer, both trilayers show an enormous AMR. Besides, $C_3R_6C_3$ exhibits almost two times larger AMR than that of $C_{12}R_6C_{12}$, indicating an enhanced magnetic anisotropy induced by "*chain-type*" SCO.

The intriguing magnetic properties of $C_3R_6C_3$ and $C_{12}R_6C_{12}$ trilayers were further investigated by Hall measurements presented in **Figure 3**. Typically, the Hall resistivity ($\rho_{yx}$) in a ferromagnet can be expressed as,[20-23]

$$\rho_{yx} = R_0 H + R_S M_z$$

where $R_0$ is the ordinary Hall effect (OHE) coefficient, $R_S$ is the anomalous Hall effect (AHE) coefficient, and $M_z$ is the out-of-plane magnetization. In order to separate the anomalous Hall resistivity from the Hall resistance, the contribution from OHE is subtracted from $\rho_{yx}$. Therefore, ($\rho_{yx} - R_0 H$) directly represents the spin-orbit coupling and the magnetization of an SRO layer. Due to the temperature dependence of band crossings near the Fermi energy and Berry phase in the momentum space, both of the temperature and magnetization contribute to anomalous Hall resistivity. Figures 3a and 3b show the field-dependent ($\rho_{yx} - R_0 H$) at various temperatures for $C_3R_6C_3$ and $C_{12}R_6C_{12}$, respectively. At a fixed temperature, a saturation ($\rho_{yx} - R_0 H$) appears above a critical magnetic field, opening a square-like hysteresis loop. The the saturation ($\rho_{yx} - R_0 H$) increases with decreasing temperature. Figure 3d shows the temperature dependent saturation ($\rho_{yx} - R_0 H$)$_{5T}$ for both trilayers. A direction comparison of *M-T* curves when the magnetic fields applied along the out-of-plane direction is plotted in Figure 3c. Strikingly, although the ($\rho_{yx} - R_0 H$)$_{5T}$ is the same for both trilayers at 5 K, ($\rho_{yx} - R_0 H$)$_{5T}$ in $C_{12}R_6C_{12}$ decays faster than that in $C_3R_6C_3$ with increasing temperature. The AHE curves nearly mimic the *M-T* trends, demonstrating the magnetic origin of AHE. Please note that the ($\rho_{yx}$-$R_0H$)-$T$ curves do not exhibit hump-like features near $H_C$, indicating that our SRO layers show neither an intrinsic topological Hall effect[21,24] nor two-channel anomalous Hall effect with opposite sign originated from sample's inhomogeneity.[33,34]

We further measured the field-dependent magnetization (*M-H*) curves for both trilayers. Both trilayers exhibit a perpendicular magnetic anisotropy (PMA) (Supporting Information, Figure S6). They reach a saturation state at perpendicular field of ~ 2 T, whereas they are not a fully saturated under in-plane magnetic fields up to 5 T. The magnetic anisotropy constant (*K*) can be determined from the area encircled by two *M-H* curves. A direct calculation gives the *K* = $1.8 \times 10^6$ erg/cm$^3$ in C$_3$R$_6$C$_3$ and *K* = $6.1 \times 10^6$ erg/cm$^3$ in C$_{12}$R$_6$C$_{12}$, respectively. **Figures 4**c and 4d show that the total saturation magnetization ($M_S$) of C$_{12}$R$_6$C$_{12}$ is three times larger than the total $M_S$ of C$_3$R$_6$C$_3$. The *M-H* curves can be well fitted by two hysteresis loops. The dashed and dotted lines in Figures 4c and 4d represent the different fractions of soft (SRO$_1$) and hard (SRO$_2$) magnetic phases in SRO layers, respectively. Therefore, we could logically divide an SRO layer into three parts—two interfacial regions (SRO$_1$) and one film bulk region (SRO$_2$), as shown in Figures 4a and 4b. There is a significant difference in $M_S$ of the soft magnetic hysteresis loops from SRO$_1$, whereas the SRO$_2$ in both trilayers shows similar $H_C$ and $M_S$. Figure 4e summaries the fitted $M_S$ of different magnetic phases in C$_3$R$_6$C$_3$ and C$_{12}$R$_6$C$_{12}$. The magnetic contribution from SRO$_1$ enhances from (60$\pm$2)% in C$_3$R$_6$C$_3$ to (88$\pm$2)% in C$_{12}$R$_6$C$_{12}$. We attribute these results to the interfacial structural modification. The SRO$_1$ is adjacent to the SCO layers and is directly affected by the bonding geometry at the interfaces. The $M_S$ of SRO$_1$ is dramatically suppressed when it connected with C-SCO, whereas the $M_S$ of SRO$_1$ keeps almost the same value (~ 1.3 $\mu_B$/Ru) as a SRO single layer when it is adjacent to P-SCO. However, SRO$_2$ is not sensitive to the interfacial modulation. Please note that the *M-H* loops from $SRO_2$ have identical $H_C$ and similar line shape to the $(\rho_{yx}-R_0 H)$-*T* curves in both trilayers. Therefore, we could identify that the $SRO_1$ with small $H_C$ and large $M_S$ is a ferromagnetic insulator with a high spin state and does not contribute to the AHE signals. In contrast, the $SRO_2$ with large $H_C$ and small $M_S$ is a ferromagnetic semiconductor with a low spin state. The itinerant electrons in $SRO_2$ lead to the observed AHE in both trilayers. This argument is supported by the insulating behavior evidenced by transport measurements (Figure 2a). The $C_{12}R_6C_{12}$ possesses a smaller portion of semiconducting $SRO_2$ layer, thus the resistivity of $C_{12}R_6C_{12}$ is larger than that of $C_3R_6C_3$. Since the thicknesses of $SRO_2$ layers in both trilayers are smaller than the critical thickness for MIT (~6 u.c.), the electronic state of $SRO_2$ stays in the insulating phase. Furthermore, we performed the polarized neutron reflectivity (PNR) measurements to identify the magnetization distribution across the SRO/SCO interfaces (Supporting Information, Figure S7).[35-39] A $[C_{12}R_6]_{15}$ superlattice with identical thickness of individual layers was used to increase the total net magnetic moment and the reliability of data fitting with 15 bilayer's repeats. PNR results indicate the SRO layers exhibit a small in-plane magnetization ~ 0.08 $\mu_B$/Ru under a magnetic field of 1 T and no magnetization is observed in SCO layers.

To illustrate the microscopic origin of large differences in the magnetization and AHE, we acquire the STEM-ABF images along [110] zone axis to observe oxygen columns and identify the octahedral distortion. **Figures 5**a and 5b shows the atomic-resolution HAADF images of SRO ultrathin layers sandwiched by 3 u.c.- and 12 u.c.-thick SCO layers, respectively. The interface roughness at the top and bottom interfaces between SRO and SCO is less than 1 u.c.. The termination layers at the interfaces are SrO and $RuO_2$ layers. The reprentative ABF images from the selected areas marked in (a) and (b) are shown in the insets of Figures 5c and 5d, respectively. The atomic positions of Ru and O atoms are clearly visible in dark contrast with sub-Å precision. For SRO sandwiched between C-SCO layers, the octahedral tilt is highly suppressed, and the Ru-O-Ru bonding angle is nearly 180º (Figure 5c). Conversely, as indicated in Figure 5d, the Ru-O-Ru bonding angles in $C_{12}R_6C_{12}$ reduces to $173º \pm 0.3º$. The sharp discrepancy in octahedral distortion between two cases can be attributed to the strong modulation of structural parameters. Although all layers are coherently strained to STO substrates, e.g. subject to the same misfit strain, the out-of-plane lattice constants of SRO are ~ $3.96 \pm 0.03$ Å and ~$3.90 \pm 0.02$ Å in $C_3R_6C_3$ and $C_{12}R_6C_{12}$, respectively. The shrink of out-of-plane lattice constants propagates from the adjacent SCO layers into the SRO layers. In $C_3R_6C_3$ with C-SCO, SRO could keep its $RuO_6$ octahedra without structural deformation. However, SRO would undergo a significant structural distortion to overcome the apical oxygen vacancies in the P-SCO, hindering its elastic deformation under the substrate-induced epitaxial strain. We calculate the tetragonality $c/a$ is $1.014 \pm 0.007$ and $0.998 \pm 0.005$ for SRO in $C_3R_6C_3$ and $C_{12}R_6C_{12}$, respectively. Under the same misfit strain, SRO in $C_3R_6C_3$ is more tetragonally distorted than SRO in $C_{12}R_6C_{12}$. Earlier work on the He-implantation into an SRO single film had demonstrated that a bulk-like orthorhombically distorted phase can be shift to a tetragonal structure by elongating uniaxially along the out-of-plane direction.[40] The dramatic phase transition is accompanied by a octahedral rotation pattern change from $a^-a^+c^-$ into $a^0a^0c^-$ when $c/a$ increases. Similarly, SRO single films grown on different substrates suffer the lattice-mismatch strain. The octahedral tilt preferentially occurs in the tensile-strained SRO ($c/a < 1$) layers.[41] These results are consistent with our experimental observations.

The microstructural distortion provides a solid evidence in the observed magnetization contrast in the SRO layers adjacent to different-type of SCO layers. In the SRO with tilted octahedra, four electrons occupy the $t_{2g}$ orbitals with lowest energy cost, exhibiting a high spin state. As the Ru-O-Ru bond angle is flattened, a small splitting between the $d_{xy}$ and $d_{xz}$ orbitals appears, resulting in one itinerant electron between two nearby orbitals. According to the Hund's Rules, the electron would flip its sign when it stays in the $d_{xz}$ orbital. The change of electronic states leads to a reduced saturation magnetization of SRO. The proposed scenario agrees with our experimental results that the magnetization of $C_3R_6C_3$ is smaller than that of $C_{12}R_6C_{12}$. Our results consistently agree with a previous theoretical prediction by Herklotz et al.[42] that the suppression of octahedral tilt quenches the magnetic moment of SRO. A recent experimental work by Jeong et al.,[26] in which they perform systematic theoretical calculations and XLD measurements at the Ru $L$-edges. They conclude that the unoccupied density of state of $d_{xy}$ orbital increases with increasing Ru-O-Ru bond angle. Therefore, a tetragonal-like SRO possesses a lower magnetization. These results consistently support the spin state of Ru is effectively controlled by crystallographic parameters which are connected to the adjacent oxygen coordination.

In summary, we report the structural modification of SRO layers in proximity to SCO with infinite-layer structure. The octahedral tilt in the SRO is strongly associated with oxygen coordination of SCO. When the $CuO_2$ infinite layers have a "*chain-type*" structure, the adjacent SRO exhibits a large magnetic anisotropy and anomalous Hall resistance. The opposed results are observed in the SRO layer connecting to the SCO has "*planar-type*" $CuO_2$ infinite layers. The selective manipulation of octahedral distortion in the functional oxides provides an effective means to engineer the local structural parameters which link to their transport and magnetic properties. Especially, our results establish a methodology of propagation the out-of-plane lattice constants from the structural dissimilar materials into the functional oxides with a fixed in-plane misfit strain. This approach adds another tuning knob to fine-control the interplay between competing electronic and magnetic order parameters in the artificial oxide heterostructures.

**Experimental Section**

***Sample synthesis and basic characterizations***. Trilayers and superlattices consisting of SrRuO$_3$ (SRO) and SrCuO$_2$ (SCO) were prepared on (001)-oriented SrTiO$_3$ (STO) substrates by pulsed laser deposition (PLD) using a XeCl excimer laser ($\lambda$ = 308 nm) with a fluence of 1.5 J/cm$^2$ and a frequency of 3 Hz. During the deposition, the substrates were kept at 700 ºC and the oxygen partial pressure were maintained at 100mTorr. Samples were cooled down to room temperature under pure oxygen at 100 Torr. The layer thickness was controlled by numbers of laser pulses. The thickness of SRO layer is fixed to 6 u.c., which is slightly above the critical thickness for the electronic phase transition. The SCO layers were grown with thickness of 3 and 12 u.c. in the trilayers and superlattices. The magnetic and transport properties were measured using PPMS and SQUID magnetometer on the SCO/SRO/SCO (C$_n$R$_6$C$_n$) trilayers, where *n (*= 3 and 12*)* and 6 represent the numbers of u.c. for SCO and SRO, respectively. Standard van der Pauw method was used in the resistivity and Hall measurements. Both in-plane and out-of-plane magnetic fields were applied to the samples. The [C$_n$R$_6$]$_{15}$ superlattices were used for PNR and STEM characterizations.

***STEM characterizations***. The C$_n$R$_6$C$_n$ trilayers and [C$_n$R$_6$]$_{15}$ superlattices were examined by STEM. Samples were prepared using ion milling after the mechanical thinning. HAADF and ABF imaging were performed simultaneously. The trilayer samples were measured along the pseudocubic [100] zone axis using a Nion UltraSTEM200 at Wuhan University of Technology (WUT). The [C$_n$R$_6$]$_{15}$ superlattices were investigated along the pseudocubic [110] zone axis using JEM ARM 200CF (Jeol Ltd., Japan) microscopy at Institute of Physics (IOP) of Chinese Academy of Sciences (CAS). The STEM data were analyzed by Gatan Digital Micrograph software. The lattice parameters were extracted by measuring the difference between the atomic positions of A-site elements. The error bars represent the standard deviation.

***X-ray spectroscopic measurements***. Room-temperature XAS measurements were conducted in the total electron yield (TEY) mode for both Cu *L*-edges and O *K*-edge at the beamline 4B9B of the Beijing Synchrotron Radiation Facility (BSRF). The sample's scattering plane was rotated at the angle of 30º and 90º with respect to the direction of incident x-ray beam. When the x-ray beam was perpendicular to the samples' surface, the in-plane orbital information was obtained ($I_{ab} = I_{90º}$). The out-of-plane orbital information can be calculated by $I_c = (I_{90º} - I_{30º} \cdot \sin^2 30º)/\cos^2 30º$. Thus, we could anticipate that the x-ray linear dichroism (XLD = $I_c - I_{ab}$) directly reflects the electronic states of the trilayers and superlattices.

***PNR measurements***. PNR measurements on the $[C_{12}R_6]_{15}$ superlattice were performed on both Beamline 4A at Spallation Neutron Source (SNS) of Oak Ridge National Laboratory (ORNL) and Multipurpose Reflectometer (MR) beamline at the Chinese Spallation Neutron Source (CSNS). The sample was cooled down to 10 K under an in-plane magnetic field of 1 T. The spin-polarization dependent specular reflectivity was measured as a function of the wave vector transfer (q) along the film surface normal. $R^+$ and $R^-$ are the reflectivities from the spin-up and spin-down polarized neutrons, respectively. Please note that PNR is only sensitive to the net in-plane component of the macroscopic magnetization within the layers. Therefore, the magnetization derived from PNR at 1 T is significantly smaller than the saturation magnetization of the SRO layers. PNR data were fitted using a chemical depth profile obtained from XRR fitting. We analyze PNR and XRR data using GenX software.

# FIGURES

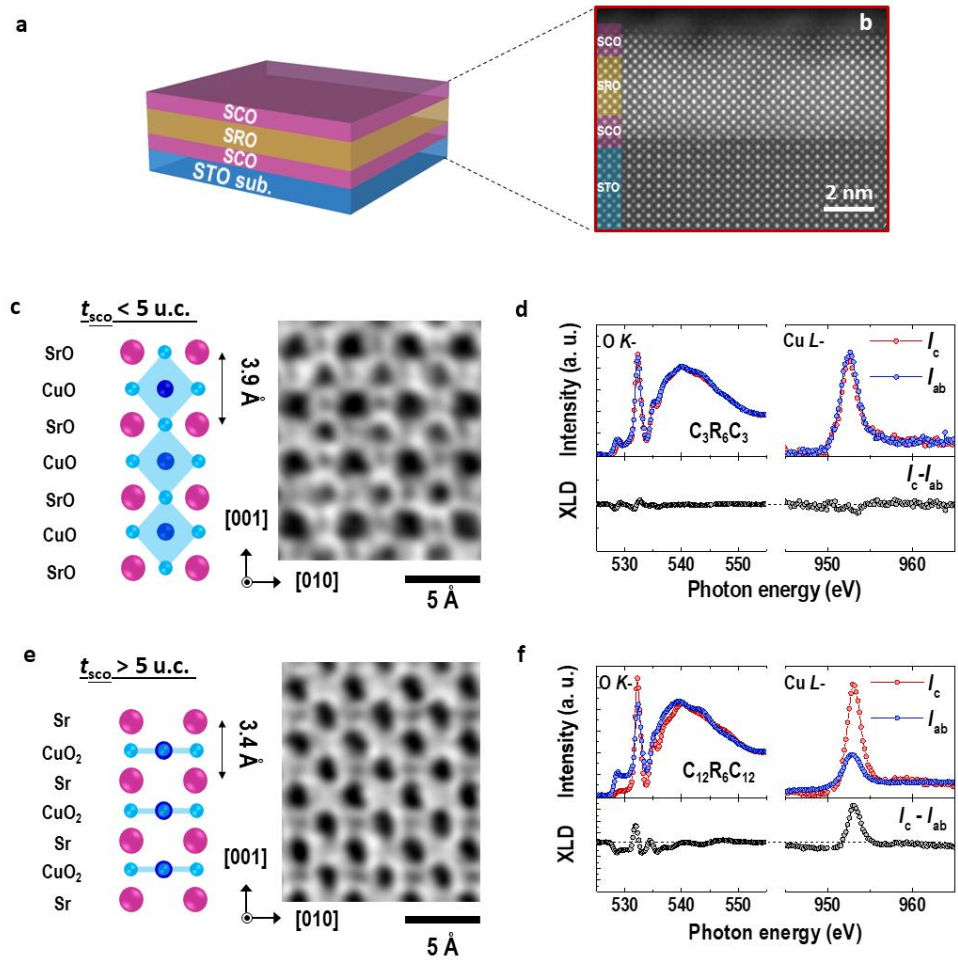

**Figure 1. Dimensional control of oxygen coordination and electronic state in SrCuO$_2$/SrRuO$_3$/SrCuO$_2$ (SCO/SRO/SCO, C$_n$R$_m$C$_n$) trilayers**, where *n* and *m* represent the number of unit cells of SCO and SRO layers, respectively. (a) Schematic illustration for a C$_n$R$_m$C$_n$ trilayer grown on STO substrate. (b) Cross-sectional HAADF-STEM image of a representative C$_3$R$_6$C$_3$ trilayer. (c) and (e) Schematic illustrations for the oxygen coordination of SrCuO$_2$ with chain-type ($t_{SCO}$ < 5 u.c.) and planar-type ($t_{SCO}$ > 5 u.c.) structures, respectively. ABF-STEM images shown on the right side are the representative SCO layers in the trilayers. These results indicate the change of oxygen coordination with increasing SCO thickness. All STEM images were taken along the pseudocubic [100] zone axis. (d) and (f) Polarization dependent XAS of O *K*-edge and Cu *L$_2$*-edge for C$_3$R$_6$C$_3$ and C$_{12}$R$_6$C$_{12}$ trilayers, respectively. The bottom panels in (d) and (f) show the x-ray linear dichroism (XLD) calculated from ($I_c$–$I_{ab}$), demonstrating a distinct orbital polarization in C$_n$R$_m$C$_n$ trilayers with different oxygen coordination.

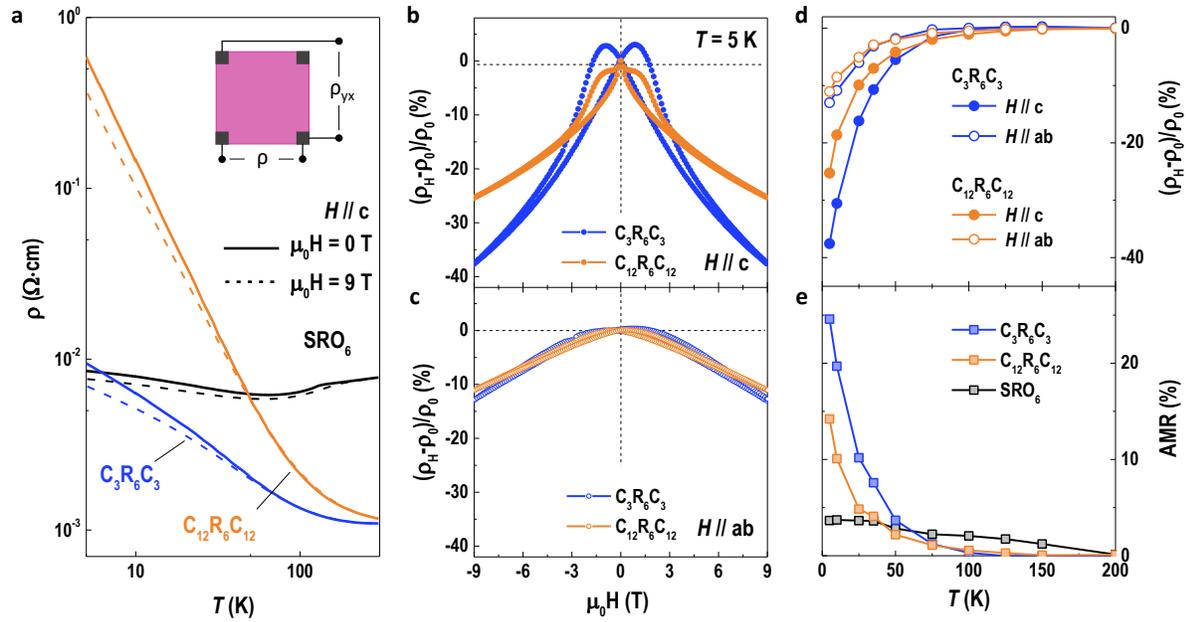

**Figure 2. Magnetotransport properties of $C_nR_6C_n$ trilayers.** (a) Temperature dependent resistivity ($\rho$) of a 6 u.c.-thick SRO single layer ($SRO_6$), $C_3R_6C_3$ and $C_{12}R_6C_{12}$ trilayers. Solid and dashed lines represent the $\rho$-$T$ curves measured at 0 and 9 T, respectively. Inset shows the geometry of electrical measurements. Magnetic field dependent magnetoresistances [MR = ($\rho_H$-$\rho_0$)/$\rho_0$] were measured at 5 K for $C_3R_6C_3$ and $C_{12}R_6C_{12}$ trilayers with magnetic field applied parallel to (b) [001] ($H$//c) and (c) [100] ($H$//ab). (d) Temperature dependent MR (top) and anisotropic magnetoresistance (bottom) (AMR = $MR_c$−$MR_{ab}$) of $SRO_6$, $C_3R_6C_3$ and $C_{12}R_6C_{12}$ trilayers.

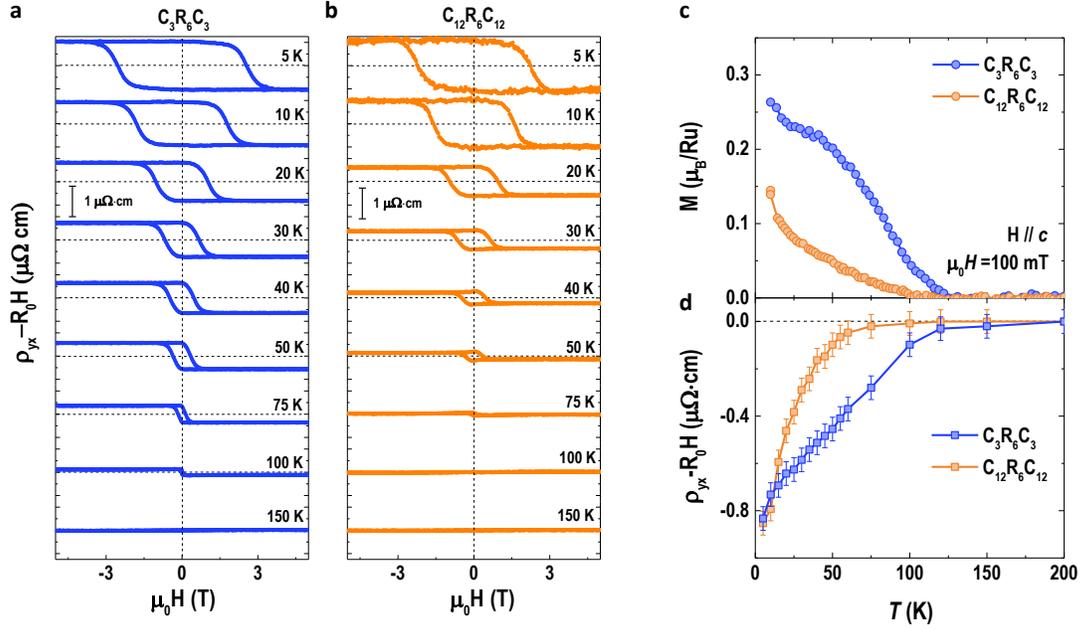

**Figure 3. Anomalous Hall resistance of $C_nR_6C_n$ trilayers.** (a) Magnetic field dependent Hall resistance ($\rho_{yx}-R_0H$) of $C_3R_6C_3$ and $C_{12}R_6C_{12}$ trilayers at various temperatures. $R_0H$ represents the ordinary Hall term which is subtracted from $\rho_{yx}$ by linear fitting in the high magnetic field region. ($\rho_{yx}-R_0H$) at each temperature is shifted for clarification. The scale bar of 1 $\mu\Omega\cdot$cm is included. (c) Temperature dependent $M$ of $C_3R_6C_3$ and $C_{12}R_6C_{12}$ trilayers. $M$ were recorded after field cooling in 100 mT applied along the out-of-plane ($H//c$) direction. (d) Temperature dependent ($\rho_{yx}-R_0H$) of $C_3R_6C_3$ and $C_{12}R_6C_{12}$ trilayers at $\mu_0H$ = 5 T.

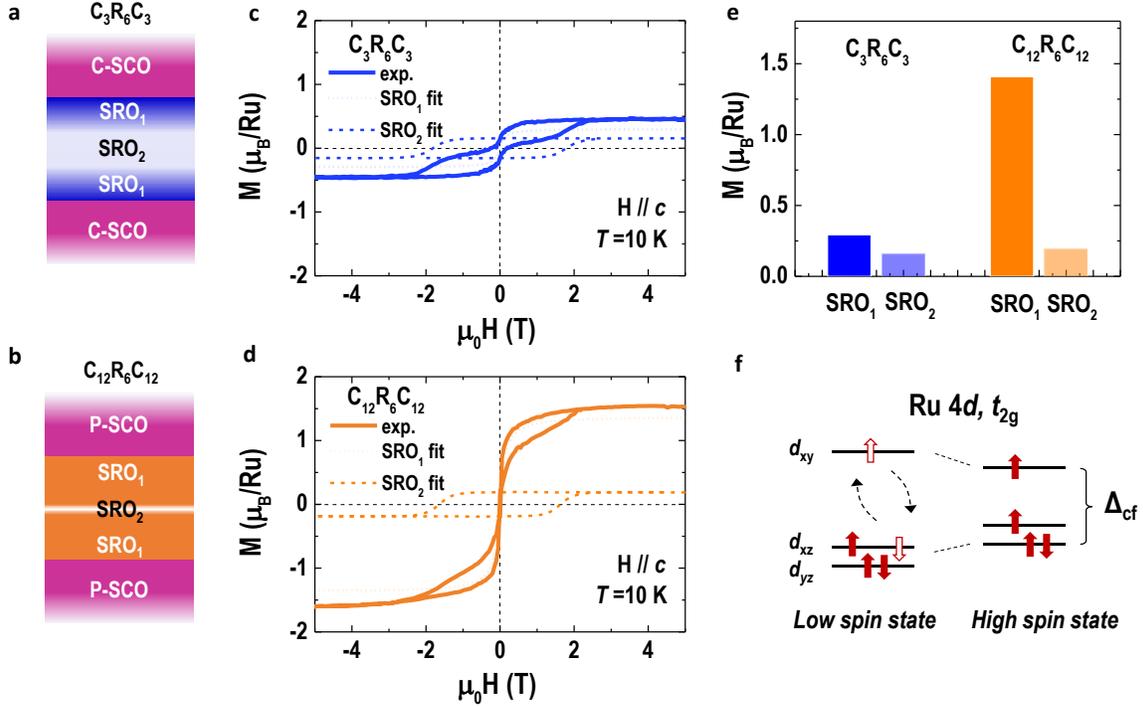

**Figure 4. Magnetic properties of $C_nR_6C_n$ trilayers**. (a) and (b) Sample geometries of $C_3R_6C_3$ and $C_{12}R_6C_{12}$ trilayers, respectively. The SRO layers were devided into two parts, $SRO_1$ and $SRO_2$. C-SCO and P-SCO represent the *chain-type* SCO and *planar-type* SCO, respectively. (c) and (d) Field dependent $M$ measured at 10 K for $C_3R_6C_3$ and $C_{12}R_6C_{12}$ trilayers, respectively. The *M-H* curves can be fitted by two hysteresis loops contributed from $SRO_1$ and $SRO_2$, respectively. The *M-H* curves from $SRO_2$ are normalized to the field-dependent anomalous Hall resistivity curves shown in Figure 3. (e) Summary of $M_S$ contributed from $SRO_1$ and $SRO_2$ in $C_3R_6C_3$ and $C_{12}R_6C_{12}$ trilayers. The interfacial $SRO_1$ in two trilayers exhibits a large difference in $M_S$, whereas the film bulk $SRO_2$ shows almost identical $M_S$. (f) Electronic structure of Ru $t_{2g}$ orbital states with low and high spin states. Red solid (empty) arrows represent fully (partially) occupied spin states, which is controlled by the crystal-field energy ($\Delta_{cf}$), *i.e.* the octahedral tilt angle.

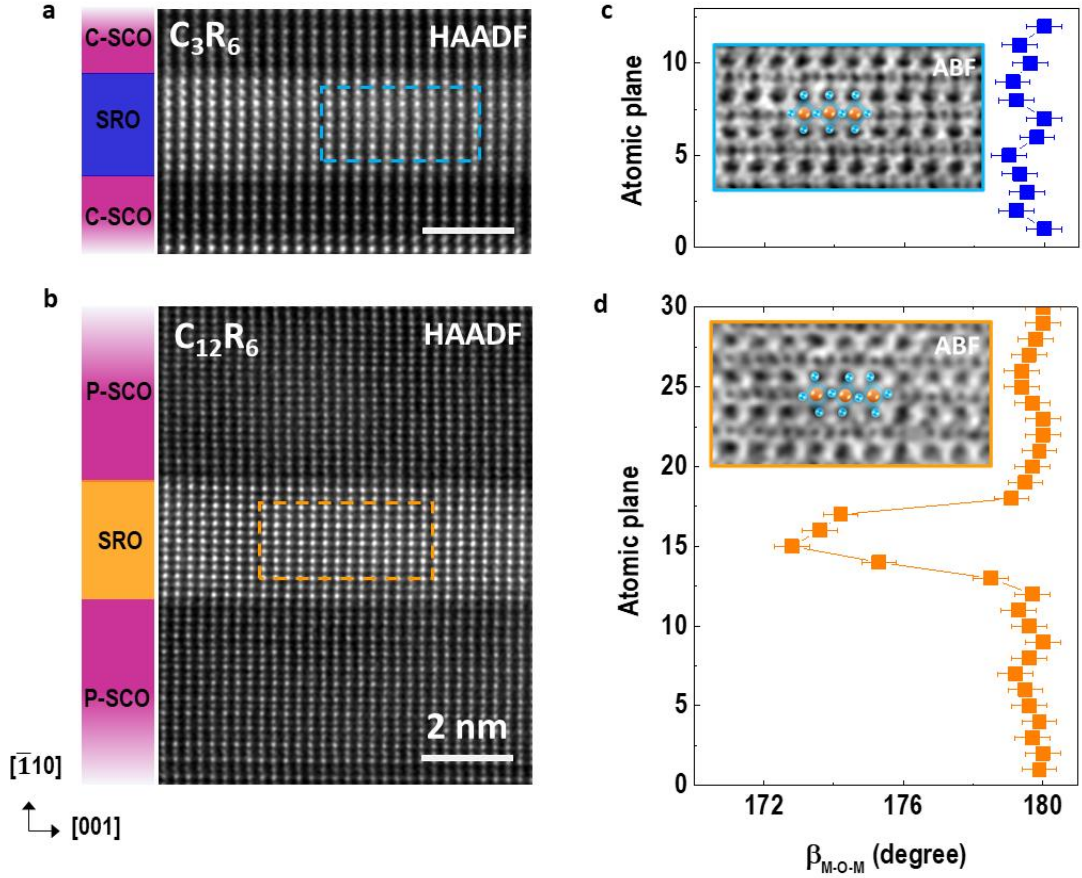

**Figure 5. Microstructure characterizations of SRO ultrathin layers sandwiched between two SCO layers.** HAADF-STEM images of representative SRO layers sandwiched between (a) 3 u.c. SCO layers and (b) 12 u.c. SCO layers. Samples were imaged along the pseudocubic [110] zone axis in the cross-sectional view. The representative ABF-STEM images from the selected colored dashed regions are shown in the insets of (c) and (d), respectively. The orange and blue spheres in the ABF-STEM images represent the Ru and O atoms, respectively. (c) and (d) The layer-position-dependent bonding angles ($\beta_{M-O-M}$) across the interfaces in $C_3R_6$ and $C_{12}R_6$ samples, respectively, where M represents transition metal ions (Ru and Cu). The error bars represent one standard deviation.

## ASSOCIATED CONTENT

### Supporting Information

Supporting Information is available online or from the author.

## AUTHOR INFORMATION

### Corresponding Author

*E-mail: kjjin@iphy.ac.cn

*E-mail: ejguo@iphy.ac.cn

### Author Contributions

These samples were grown by S. L. under the guidance of E.J.G.; TEM lamellas were fabricated with FIB milling and TEM experiments were performed by Q.H.Z., X. H. S. and L. G.; XAS measurements were conducted by S. L., J. Z., Q. J., S. C., S. C., and J. O. W.; Magnetic measurements were performed by S. L., H. G.; PNR measurements were performed by A. H., M. R. F., and T. Z. E.J.G. initiated the research. S. L. and E. J. G. wrote the manuscript with help from K. J. J. All authors participated in the discussion of manuscript.

### Notes

The authors declare no competing financial interest.

## ACKNOWLEDGMENT

This work was supported by the National Key Basic Research Program of China (Grant Nos. 2020YFA0309100 and 2019YFA0308500), the National Natural Science Foundation of China (Grant Nos. 11974390, 51902237, 52025025, and 52072400), the Beijing Nova Program of

Science and Technology (Grant No. Z191100001119112), the Beijing Natural Science Foundation (Grant No. 2202060), the program for the Innovation Team of Science and Technology in University of Henan (No. 20IRTSTHN014), and the Strategic Priority Research Program (B) of the Chinese Academy of Sciences (Grant No. XDB33030200). The XAS and XLD experiments at the beam line 4B9B of the Beijing Synchrotron Radiation Facility (BSRF) of the Institute of High Energy Physics, Chinese Academy of Sciences were conducted via a user proposal. PNR measurements at the Chinese Spallation Neutron Source (CSNS) were conducted via a user proposal (P2018121100016). The research at ORNL's SNS (PNR measurements), which is a U.S. Department of Energy (DOE), Office of Science (OS), Basic Energy Sciences (BES) scientific user facility, was conducted via a user proposal.

**REFERENCES**

(1) J. M. Rondinelli, et al., *Control of octahedral connectivity in perovskite oxide heterostructures: An emerging route to multifunctional materials discovery*, MRS Bull. **2012**, 37, 261-270.

(2) E. J. Moon, et al., *Spatial control of functional properties via octahedral modulations in complex oxide superlattices*, Nat. Commun. **2014**, 5, 5710.

(3) M. Imada, A. Fujimori, Y. Tokura, *Metal-insulator transitions*, *Rev. Mod. Phys.* **1998**, *70*, 1039.

(4) Pavlo Zubko, Stefano Gariglio, Marc Gabay, Philippe Ghosez, and Jean-Marc Triscone, *Interface Physics in Complex Oxide Heterostructures*, Annual Review of Condensed Matter Physics, **2011**, 2, 141-165.

(5) D. G. Schlom, et al., *Strain Tuning of Ferroelectric Thin Films*, Annu. Rev. Mater. Res. **2007**, 37, 589-626.

(6) J. P. Locquet et al., *Doubling the critical temperature of $La_{1.9}Sr_{0.1}CuO_4$ using epitaxial strain*, Nature **1998**, 394, 453.

(7) H. N. Lee et al., *Strong polarization enhancement in asymmetric three-component ferroelectric superlattices*, Nature **2005**, 433, 395.

(8) J. H. Haeni, et al., *Room-temperature ferroelectricity in strained $SrTiO_3$*, Nature **2004**, 430, 758.

(9) A. V. Boris, et al., *Dimensionality Control of Electronic Phase Transitions in Nickel-Oxide Superlattices*, Science **2011**, 332, 937.

(10) Q. Gan, R. A. Rao, C. B. Eom, J. L. Garrett, M. Lee, *Direct measurement of strain effects on magnetic and electrical properties of epitaxial $SrRuO_3$ thin films*, Appl. Phys. Lett. **1998**, *72*, 978.

(11) G. Koster, L. Klein, W. Siemons, G. Rijnders, J. S. Dodge, C. B. Eom, D. H. A. Blank, M. R. Beasley, *Structure, physical properties, and applications of $SrRuO_3$ thin films*, Rev. Mod. Phys. **2012**, *84*, 253.

(12) D. Kan et al., *Strain Effect on Structural Transition in $SrRuO_3$ Epitaxial Thin Films*, Cryst. Growth Des. **2013**, 23, 1129-1136.

(13) J. Xia, W. Siemons, G. Koster, M. R. Beasley, A. Kapitulnik, Critical thickness for itinerant ferromagnetism in ultrathin films of $SrRuO_3$, Phys. Rev. B **2009**, *79*, 140407.

(14) R. Aso, et al., *Strong Dependence of Oxygen Octahedral Distortions in SrRuO$_3$ Films on Types of Substrate-Induced Epitaxial Strain*, Cryst. Growth Des. **2014**, 14, 6478-6485.

(15) Y. Kats, I. Genish, L. Klein, J. W. Reiner, M. R. Beasley, *Large anisotropy in the paramagnetic susceptibility of SrRuO$_3$ films*, Phys. Rev. B **2005**, *71*, 100403.

(16) D. Toyota, I. Ohkubo, H. Kumigashira, M. Oshima, T. Ohnishi, M. Lippmaa, M. Takizawa, A. Fujimori, K. Ono, M. Kawasaki, H. Koinuma, *Thickness-dependent electronic structure of ultrathin SrRuO3 films studied by in situ photoemission spectroscopy*, Appl. Phys. Lett. **2005**, *87*, 162508.

(17) D. Kan, et al., *Tuning magnetic anisotropy by interfacially engineering the oxygen coordination environment in a transition metal oxide*, Nature Materials **2016**, 15, 432.

(18) S. Thomas, et al., *Localized Control of Curie Temperature in Perovskite Oxide Film by Capping-Layer-Induced Octahedral Distortion*, Phys. Rev. Lett. **2017**, 119, 177203.

(19) S. Lin, et al., *Switching Magnetic Anisotropy of SrRuO$_3$ by Capping-Layer-Induced Octahedral Distortion*, Phys. Rev. Appl. **2020**, 13, 034033.

(20) Z. Fang, N. Nagaosa, K. S. Takahashi, A. Asamitsu, R. Mathieu, T. Ogasawara, H. Yamada, M. Kawasaki, Y. Tokura, and K. Terakura, *Anomalous Hall Effect and Magnetic Monopoles in Momentum-Space*, Science **2003**, *302*, 92.

(21) J. Matsuno, N. Ogawa, K. Yasuda, F. Kagawa, W. Koshibae, N. Nagaosa, Y. Tokura, M. Kawasaki, *Interface-driven topological Hall effect in SrRuO$_3$-SrIrO$_3$ bilayer*, Sci. Adv. **2016**, *2*, e1600304.

(22) Q. Qin, *et al.*, *Emergence of Topological Hall Effect in a SrRuO$_3$ Single Layer*, Adv. Mater. **2019**, 31, 1807008.

(23) Y. Gu, Y.-W. Wei, K. Xu, H. Zhang, F. Wang, F. Li, M. S. Saleem, C.-Z. Chang, J. Sun, C. Song, J. Feng, X. Zhong, W. Liu, Z. Zhang, J. Zhu, F. Pan, *Interfacial oxygen-octahedral-tilting-driven electrically tunable topological Hall effect in ultrathin SrRuO$_3$ films*, J. Phys. D: Appl. Phys. **2019**, *52*,404001.

(24) L. Wang, Q. Feng, Y. Kim, R. Kim, K. H. Lee, S. D. Pollard, Y. J. Shin, H. Zhou, W. Peng, D. Lee, W. Meng, H. Yang, J. H. Han, M. Kim, Q. Lu, T. W. Noh, *Ferroelectrically tunable magnetic skyrmions in ultrathin oxide heterostructures*, Nature Materials **2018**, 17, 1087–1094.

(25) Z. Cui, A. J. Grutter, H. Zhou, H. Cao, Y. Dong, D. A. Gilbert, J. Wang, Y. S. Liu, J. Ma, Z. Hu, J. Guo, J. Xia, B. J. Kirby, P. Shafer, E. Arenholz, H. Chen, X. Zhai, Y. Lu, *Correlation-driven eightfold magnetic anisotropy in a two-dimensional oxide monolayer*, Sci. Adv. **2020**, *6*, eaay0114.

(26) S. G. Jeong, et al., *Propagation Control of Octahedral Tilt in SrRuO$_3$ via Artificial Heterostructuring*, Adv. Sci. **2020**, 2001643.

(27) H. Boschker, et al., *Ferromagnetism and Conductivity in Atomically Thin SrRuO$_3$*, Phys. Rev. X **2019**, 9. 011027.

(28) Seung Gyo Jeong, et al., *Phase Instability amid Dimensional Crossover in Artificial Oxide Crystal*, Phys. Rev. Lett. **2020**, 124, 026401.

(29) I. A. Zaliznyak, C. Broholm, M. Kibune, M. Nohara, H. Takagi, *Anisotropic Spin Freezing in the S=1/2 Zigzag Chain Compound SrCuO$_2$*, Phys. Rev. Lett. **1999**, *83*, 5370.

(30) Z. C. Zhong, G. Koster, P. J. Kelly, *Prediction of thickness limits of ideal polar ultrathin films*, Phys. Rev. B **2012**, *85*, 121411.

(31) D. Samal, H. Y. Tan, H. Molegraaf, B. Kuiper, W. Siemons, S. Bals, J. Verbeeck, G. V. Tendeloo, Y. Takamura, E. Arenholz, C. A. Jemkins, G. Rijnders, G. Koster, *Experimental Evidence for Oxygen Sublattice Control in Polar Infinite Layer $SrCuO_2$*, Phys. Rev. Lett. **2013**, *111*, 096102.

(32) Z. Liao, E. Skoropata, J. W. Freeland, E. J. Guo, R. Desautels, X. Gao, C. Sohn, A. Rastogi, T. Z. Ward, T. Zou, T. Charlton, M. R. Fitzsimmons, and H. N. Lee, *Large orbital polarization in nickelate-cuprate heterostructures by dimensional control of oxygen coordination*, Nat. Commun. **2019**, *10*, 589.

(33) G. Kimbell, P. M. Sass, B. Woltjes, E. K. Ko, T. W. Noh, W. Wu, and Jason W. A. Robinson, *Two-channel anomalous Hall effect in $SrRuO_3$*, Phys. Rev. Mater. **2020**, 4, 054414.

(34) L. Wu, F. Wen, Y. Fu, J. H. Wilson, X. Liu, Y. Zhang, D. M. Vasiukov, M. S. Kareev, J. H. Pixley, and J. Chakhalian, *Berry phase manipulation in ultrathin $SrRuO_3$ Films*, Phys. Rev. B **2020**, 102, 220406(R).

(35) J. Chakhalian, J. W. Freeland, H.-U. Habermeier, G. Cristiani, G. Khaliullin, M. van Veenendaal, and B. Keimer, *Orbital reconstruction and covalent bonding at an oxide interface*, Science **2007**, 318, 1114-1117.

(36) J. Chakhalian, J. W. Freeland, G. Srajer, J. Strempfer, G. Khaliullin, J. C. Cezar, T. Charlton, R. Dalgliesh, C. Bernhard, G. Cristiani, H.-U. Habermeier, and B. Keimer, *Magnetism at the interface between ferromagnetic and superconducting oxides*, Nat. Phys. **2006**, 2, 244-248.

(37) M. R. Fitzsimmons, S. D. Bader, J. A. Borchers, G. P. Felcher, J. K. Furdyna, A. Hoffmann, J. B. Kortright, I. K. Schuller, T. C. Schulthess, S. K. Sinha, M. F. Toney, D. Weller, S. Wolf, *Neutron scattering studies of nanomagnetism and artificially structured materials*, J. Magn. Magn. Mater. **2004**, 271, 103.

(38) S. Singh, J. T. Haraldsen, J. Xiong, E. M. Choi, P. Lu, D. Yi, X.-D. Wen, J. Liu, H. Wang, Z. Bi, P. Yu, M. R. Fitzsimmons, J. L. MacManus-Driscoll, R. Ramesh, A. V. Balatsky, J. X. Zhu, Q. X. Jia, *Induced magnetization in $La_{0.7}Sr_{0.3}MnO_3/BiFeO_3$ superlattices*, Phys. Rev. Lett. **2014**, 113, 047204.

(39) E. J. Guo, J. R. Petrie, M. A. Roldan, Q. Li, R. D. Desautels, T. Charlton, A. Herklotz, J. Nichols, J. van Lierop, J. W. Freeland, S. V. Kalinin, H. N. Lee, M. R. Fitzsimmons, *Spatially Resolved Large Magnetization in Ultrathin $BiFeO_3$*, Adv. Mater. **2017**, *29*, 1700790.

(40) A. Herklotz, A. T. Wong, T. Meyer, M. D. Biegalski, H. N. Lee, and T. Z. Ward, *Controlling Octahedral Rotations in a Perovskite via Strain Doping*, Sci. Rep. **2016**, 6, 26491.

(41) R. Aso, D. Kan, Y. Shimakawa, H. Kurata, *Control of Structural Distortions in Transition‐Metal Oxide Films through Oxygen Displacement at the Heterointerface*, Adv. Funct. Mater. **2014**, 24, 5177-5184.

(42) A. Herklotz and K. Doerr, *Characterization of tetragonal phases of $SrRuO_3$ under epitaxial strain by density functional theory*, Eur. Phys. J. B. **2015**, 88, 60.